\documentclass[bib]{statapress}

\usepackage[crop,newcenter,frame]{pagedims}

\usepackage{sj}

\usepackage{epsfig}

\usepackage{stata}

\usepackage{amsmath}

\usepackage{shadow}

\sjsetissue{$vv$}{$ii$}{$mm$}{$yyyy$}

\usepackage{threeparttable} 
\usepackage{tabularx}

\usepackage{booktabs}
\usepackage{graphicx}
\usepackage{wrapfig}
\usepackage{etoolbox}

\begin{document}

\newcommand\tblue[1]{{\color{blue}#1}}  
\newcommand\tgreen[1]{{\color{green}#1}}  
\newcommand\tred[1]{{\color{red}#1}}  
\newbool{grayscale}
\setbool{grayscale}{false}   

\newcommand{\figfile}[1]{%
  \ifbool{grayscale}{#1_gs.png}{#1.png}%
}

\inserttype[st0001]{article}
\author{Schonlau and Yang}{%
  Matthias Schonlau\\University of Waterloo\\Waterloo, Canada\\schonlau@uwaterloo.ca
  \and
  Tiancheng Yang \\ University of Waterloo\\Waterloo, Canada\\ t77yang@uwaterloo.ca
}
\title[The hammock plot]{The Hammock Plot: Where Categorical and Numerical Data Relax Together}
\maketitle 
\setcounter{tocdepth}{3} 

\begin{abstract}
Effective methods for visualizing data involving multiple variables, including categorical ones, are limited.
The hammock plot \citep{schonlau2003visualizing} visualizes both categorical and numerical variables using parallel coordinates.  
We introduce the Stata implementation  \texttt{hammock}. 
We give numerous examples that explore highlighting, missing values, putting axes on the same scale, and tracing an observation across variables. Further, we discuss  parallel univariate plots as an edge case of hammock plots. We also present and make publicly available a new dataset on the 2020 Tour de France. 

\keywords{\inserttag, visualization, hammock plot, clustergram, parallel coordinate plot, alluvial plot, Sankey diagram}
\end{abstract}


\section{Introduction}
Leonardo da Vinci wrote, a poet would struggle to  ``describe with words what a painter is able to [depict] in an instant" \citep{janson2004history}.  But not everybody can paint like da Vinci. Likewise, it is often easier to write a few pages of text than to construct a sensible graph. Arguably, readers appreciate graphs more than text. Empirical evidence suggests  graphs even lend an aura of scientific credibility \citep{tal2015looks}.  
For those of us who cannot paint like da Vinci, software makes it easier to create graphs. In this paper we introduce a Stata command for creating hammock plots.

For visualizing one or two variables there are many choices: bar charts, stacked and grouped bar charts, scatter plots, pie charts and others. A scatter plot matrix and a parallel coordinate plot are effective for visualizing more than two continuous variables. Mosaic plots \citep{hofmann2008mosaic} are suitable for more than two categorical variables, though they are harder to interpret. Another option is to show multiple plots that are linked through highlighting. For displaying multiple variables that contain a mixture of categorical or numerical variables in a single plot, we need a hammock plot or one of its variations. This paper showcases the hammock plot and its Stata implementation.

To introduce a less familiar plot, it may be easiest to give an example using a data set that many Stata users are familiar with, the auto data. In this data set, each of the 74 observations represent one car.  Each car's repair record was rated with 1-5 stars, 5 being best.
Figure~\ref{f:scatterhammock} shows two plots: 
\begin{figure}[htb]
\centering
\hspace*{0cm}
\includegraphics[scale=0.4]{\figfile{stata_output/scatterhammock_annotated}}
\caption{A scatter plot and a corresponding hammock plot of repair record (1-5 ``stars'') vs car origin from the auto data. The hammock plot has parallel axes. The height of a univariate bar (on an axis)  is proportional to frequency of the corresponding category. The width of a  connector (between two axes) is proportional to the corresponding bivariate frequencies. The frequency of missing connectors (e.g. repair record ``1'' to ``foreign'') is zero.  With only two axes, the usefulness of the hammock plot may not yet be apparent.}
\label{f:scatterhammock}
\end{figure}
a scatter plot of the variables repair record and car origin (foreign vs domestic) and a corresponding hammock plot. The repair record takes the values 1 through 5 corresponding to 1 through 5 ``stars''. To reduce overplotting, the scatter plot uses jittering, i.e., we add a little bit of random noise. In a scatter plot the two axes are aligned at a right angle. A hammock plot uses parallel coordinates: the axis for repair record is parallel to the axis for car origin. On the axes, there are univariate bars. They display  univariate statistics: the height of the bar is proportional to the number of observations it contains.  Axes are connected by connectors. The width of a connector is also proportional to the number of observations it contains.

The connectors represent a visualization of the two-way contingency table of repair record and car origin:

\begin{stlog}
. tab rep78 foreign, cell nofreq
{\smallskip}
    repair {\VBAR}      Car origin
    record {\VBAR}  Domestic    Foreign {\VBAR}     Total
\HLI{11}{\PLUS}\HLI{22}{\PLUS}\HLI{10}
         1 {\VBAR}      2.90       0.00 {\VBAR}      2.90 
         2 {\VBAR}     11.59       0.00 {\VBAR}     11.59 
         3 {\VBAR}     39.13       4.35 {\VBAR}     43.48 
         4 {\VBAR}     13.04      13.04 {\VBAR}     26.09 
         5 {\VBAR}      2.90      13.04 {\VBAR}     15.94 
\HLI{11}{\PLUS}\HLI{22}{\PLUS}\HLI{10}
     Total {\VBAR}     69.57      30.43 {\VBAR}    100.00 
{\smallskip}
\nullskip
\end{stlog}
There are eight connectors between the two axes and eight non-zero cells in the two-way contingency table. The  bars on the axes visualize the marginal statistics of the two variables shown in the ``Total'' row and column of the contingency table.

\begin{figure}[htbp]
\centering
\hspace*{0cm}
\includegraphics[width=1.\textwidth]{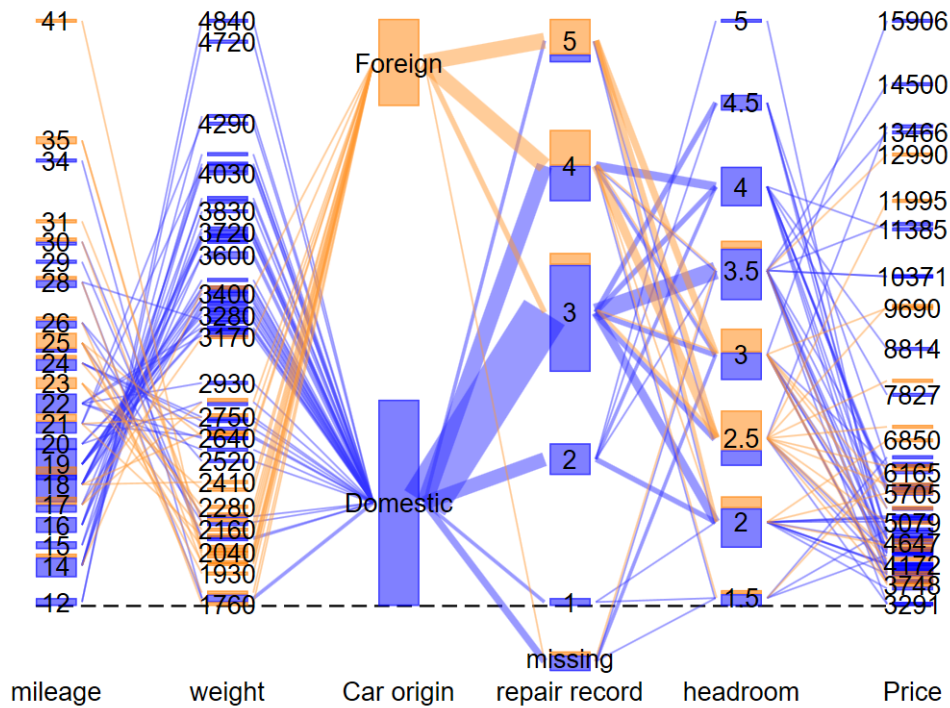}
\caption{Hammock plot of the auto data. Foreign cars are highlighted. A category for missing values, where present, is displayed near the bottom. Here, the black line was added to show the separation between the areas for missing and non-missing values.}
\label{f:auto_foreign}
\end{figure}

A hammock plot with two axes may not be particularly useful. However, parallel axes have the advantage that we can add additional axes.
Figure~\ref{f:auto_foreign} shows a hammock plot \citep{schonlau2003visualizing} of a half dozen variables of the auto data set. (We chose some variables to illustrate the hammock plot; a different set of variables could have been chosen. The Stata code for this plot is given later in Section~\ref{s:high}.)

The graph consists of axes with univariate information (univariate bars)  and connectors  that connect the axes. We discuss each in turn.

Each variable is shown on a parallel axis; the variable name is displayed beneath the axes. 
For categorical variables, each axis  contains a bar that corresponds to one category. The height of the bar is proportional to the number of observations it represents. 
For example, the variable foreign in the last axis has two categories, ``foreign'' and ``domestic'' representing foreign and domestic cars. We see there are more than twice as many domestic cars as foreign cars in this data set. 
For numerical variables, there is  also a ``bar'' for each value. However, for values occurring only once, the bar is so thin that it looks like a line. So, the univariate display for a numerical value looks like a rug plot (see the discussion in \citet{cox2025rug}), which we typically see at the bottom of a density plots to show the individual data points of a continuous variable. 
Each axis is scaled individually from minimum to maximum of the corresponding variable. For example, the axis for mileage ranges from 12 to 41; that of weight from 1760 to 4840.

``Missing values'' is  a separate category and, optionally, a space for missing values is reserved at the bottom. 
Figure~\ref{f:auto_foreign} features an optional line that separates the area between missing and non-missing values.  The variable repair record has a  bar at the bottom which indicates its missing values. 

 Neighboring axes are connected by connectors. In Figure \ref{f:auto_foreign}, the connectors are rectangles, but other shapes could be used (e.g., parallelograms, which are faster to compute).  The width of the connector is proportional  to the number of observations it contains. Connectors containing very few observations appear as a line (even though they are just thin connectors).  The connectors each represent a cell of the two-way contingency table of the two adjacent variables, i.e., they represent bivariate statistics. 

Figure \ref{f:auto_foreign} highlights foreign cars in a different color. We can follow the observations belonging to foreign cars throughout the graph. For example, we see that foreign cars tend to have 
less headroom, 
a better repair record,  lower weight, and higher mileage. 
For the continuous variables mileage (miles per gallon) and weight we can also make a statement about correlations. Mileage and weight are negatively correlated because the connectors cross in opposite directions: high values of mileage correspond to low values of weight and conversely. In fact, the correlation between mileage and weight turns out to be  $-0.8$.

The remainder of this paper is as follows. 
Section \ref{s:relatedplots} gives a brief history of the hammock plot and similar plots that followed.
Section~\ref{s:syntax} gives an overview over syntax and options.
Section~\ref{s:example} gives numerous examples and explains the Stata syntax in the context of the examples. 
Section~\ref{s:conclusion} concludes.

\section{\label{s:relatedplots}A brief history of the hammock plot and related plots}
In 2003, \citet{schonlau2003visualizing} proposed the hammock plot to visualize both categorical and numerical variables. 
The hammock plot  was inspired by the clustergram \citep{schonlau2002clustergram,schonlau2004clustergram2}, a program to visualize how clusters form as you increase the number of clusters. 
Like the parallel coordinate plot \citep{inselberg1985plane,wegman1990hyperdimensional} for continuous variables, the hammock plot also uses parallel axes. However, unlike parallel coordinate plots, hammock plots use two-dimensional plotting elements rather than lines to connect adjacent axes.  A hammock plot of only continuous variables (without labels, univariate statistics, and missing values) is a parallel coordinate plot. More recently, \citet{schonlau2024hammock} proposed connecting neighboring axes with rectangles rather than parallelograms. This avoids the so-called reverse line-width illusion.

Several hammock-type plots have been proposed including the 
 ParSets plot \citep{kosara2006parallel},
 alluvial plot \citep{rosvall2010alluvial},
 common angle plot \citep{hofmann2013common}, 
 and the cpcp plot \citep{pilhoefer2013new}.
 Such plots differ in how they connect two adjacent categorical variables (e.g., rectangle vs. rounded shapes),
 two adjacent numerical variables (e.g., rectangle vs. line),
 two adjacent numerical-categorical variables (e.g., rectangle vs. triangle), whether they allow numerical variables or force categorical encodings, and how they space 
 out numerical variables along a numerical axis.
 
 The generalized parallel coordinate plot (GPCP) \citep{vanderplas2023penguins} uses individual lines (rather than areas) to connect neighboring axes. The plot resolves ties by plotting parallel lines (rather than areas). For small data sets, this allows tracing individual observations across the plot.  To trace individual observations,
 the hammock plot uses highlighting (see Section~\ref{s:tracing}).

Perhaps the best known plot among the hammock-type plots is  the alluvial plot \citep{rosvall2010alluvial}.
 Whereas the hammock plot uses parallelograms or rectangles to connect axes, alluvial plots use rounded curves. These plots were initially only used to visualize network variables over time. Alluvial plots are now also used to visualize categorical variables; they do not accommodate numerical variables. In Stata, alluvial plots are implemented in the package \texttt{\textbf{alluvial}} \citeb{alluvial2024}.

A similar looking plot, the Sankey diagram \citep{sankey1898diagram,schmidt2008sankey},  visualizes flows of energy and materials rather than variables. 
A Sankey diagram may have circular flows or one flow may stop in the middle of the graph.
The Sankey diagram is implemented in Stata \citep{sankeystata}.  The names ``Sankey diagram'' and ``Alluvial plot'' are sometimes conflated in the literature. If a plot visualizes variables, it is not a Sankey diagram.
A more detailed overview over all of these plots can be found in \citet{schonlau2024hammock}.
A Stata implementation of the parallel coordinate plot is in \texttt{\textbf{parplot}} \citep{parplot07}.

\section{\label{s:syntax}Syntax}

\begin{verbatim}
hammock varlist [if] [in] , [options]
\end{verbatim}

The following sections list options alongside a brief description. As in the helpfile, we underline abbreviations for options. For a more detailed description please consult the helpfile.
For the most recent version of this command see the 
Boston SSC archive (\texttt{ssc install hammock}) or my github page \citep{hammocksoftware2025}. 

\subsection{Highlighting}
\begin{description}
  \item[\texttt{\underline{hivar}iable(varname)}]
  Name of the variable to highlight.

  \item[\texttt{\underline{hival}ues(string)}]
  List of values of \emph{hivariable} to highlight.

  \item[\texttt{\underline{col}orlist(str)}]
  Default color and colors used for highlighting.

  \item[\texttt{uni\_colorlist(str)}]
  Default color and colors used for highlighting univariate bars.
\end{description}

\subsection{Layout of univariate bars}
\begin{description}
  \item[\texttt{\underline{nouni}bar}]
  Do not show univariate bars.

  \item[\texttt{uni\_fraction(real)}]
  Proportion of vertical space allocated to univariate bars.
\end{description}

\subsection{Layout of labels}
\begin{description}
\item[\texttt{\underline{nolab}el}]
  Do not show value labels.

\item[\texttt{label\_methodlist(str)}]
  Specify individual labeling methods for each variable.

\item[\texttt{label\_too\_many(real)}]
  When a variable has at least \emph{label\_too\_many} unique values,
  show only the minimum and maximum label.

\item[\texttt{label\_min\_dist(real)}]
  For method \emph{min\_dist}, show only labels that are at least
  \emph{label\_min\_dist} apart.

\item[\texttt{labelopt(str)}]
  Options passed to \texttt{added\_text\_options}, for example to control
  label text size.

\item[\texttt{label\_format(str)}]
  Display format of numeric labels.
\end{description}

\subsection{Layout of labels}
\begin{description}
  \item[\texttt{\underline{nolab}el}]
  Do not show value labels.

  \item[\texttt{label\_min\_dist(real)}]
  Specify the minimum distance between two labels on the same axis.

  \item[\texttt{labelopt(str)}]
  Options passed to \texttt{added\_text\_options}, for example to control label text size.

  \item[\texttt{label\_format(str)}]
  Display format for numerical labels.
\end{description}

\subsection{Missing values}
\begin{description}
  \item[\texttt{missing}]
  Display missing values.

  \item[\texttt{\underline{m}issing\_fraction(real)}]
  Proportion of vertical space allocated to missing values.
\end{description}

\subsection{Layout of connectors between axes}
\begin{description}
  \item[\texttt{\underline{bar}width(real)}]
  Change the width of connectors to reduce clutter.

  \item[\texttt{\underline{minbar}freq(int)}]
  Specify the minimum connector width.

  \item[\texttt{shape(str)}]
  Shape of the connectors: \emph{parallelogram} (or \emph{par}) or \texttt{rectangle} (default).

  \item[\texttt{outline}]
  Outline the edges of semi-translucent connectors (rarely needed).
\end{description}

\subsection{Other options}
\begin{description}
  \item[\texttt{\underline{spa}ce(real)}]
  Control the fraction of space allocated to labels and univariate bars relative to connectors.

  \item[\texttt{\underline{same}scale(varlist)}]
  Use the same axis scale for each variable specified.

  \item[\texttt{subspace(real)}]
  Adjust empty space between univariate bars and connectors (rarely needed).

  \item[\texttt{\underline{aspect}ratio(real)}]
  Aspect ratio of the plot region (rarely needed).

  \item[\texttt{graph\_options}]
  Additional options passed through to \texttt{graph, twoway}.
\end{description}

\section{\label{s:example}Examples}
Our first example features a mixture of numerical and categorical variables, highlighting, and missing values (Section~\ref{s:high}).
In the following section, Section~\ref{s:tracing}, we trace individual observations in a dataset related to the  Tour de France. The variables in this example all use the same scale (seconds).  The \texttt{samescale} option  shows the effect of standardizing the layout across axes. 
Section~\ref{s:connectors}, contains an example where  our interest lies in connectors between axes; the univariate bars are largely redundant and are omitted.
Section~\ref{s:paruni} shows the other extreme: parallel univariate bars without connectors. The examples include a graphical representation of socio-demographic table, enriching or replacing the typical ``Table 1'' in empirical studies.

\subsection{Highlighting\label{s:high}}
The hammock plot in Figure~\ref{f:auto_foreign} highlighted foreign cars. The print version of this figure (and all figures) is in grayscale; the online version is in color.  The color version was created with the following command: 

\begin{stlog}
. sysuse auto, clear
(1978 automobile data)
{\smallskip}
. label var rep78 "repair record"
{\smallskip}
. label var mpg "mileage"
{\smallskip}
. label var weight "weight"
{\smallskip}
. label var headroom "headroom"
{\smallskip}
. 
. hammock mpg weight foreign rep78  headroom price, ///
>  missing ///
>  hivar(foreign) hival(1) ///
>  label_format(\%8.0g) ///
>  yline(8) ///
>  colorlist(ltblue sandb) 
\nullskip
\end{stlog}

The grayscale version replaces the colors using  \texttt{colorlist(gs12 gs8)}. 
The hammock plot displays variable labels (when present) and the display looks best with short labels. We therefore label some variables. 

The option  \texttt{missing} adds room at the bottom across all variables for a category for any missing values.  We highlight the variable and specific values of that variable with \texttt{hivar} and \texttt{hival}. The label on the plot is not the number highlighted (1), but its label (``foreign"). 
 Likewise, the variable name shown below each axis is the variable label. So, we see the label ``mileage" rather than the variable name ``mpg".   We can highlight (\texttt{hival(0)}) a single value, missing values (\texttt{hival(.)}), a range (\texttt{hival(>27)}) or  several values in different colors (e.g., \texttt{hival(1 4 5)}). It is also possible to specify an external variable for highlighting, i.e., a variable that  is not among the variables in the list of variables shown in the graph.
 
By default, the labels of the variable price in Figure \ref{f:auto_foreign} are written in scientific format if the numbers get too large. 
We can change this by adjusting the format in the \texttt{label\_format} option. 

\subsubsection{Specifying Colors}
\begin{figure}[htbp]
\centering
\includegraphics[width=1\textwidth, trim=10pt 10pt 10pt 10pt, clip]{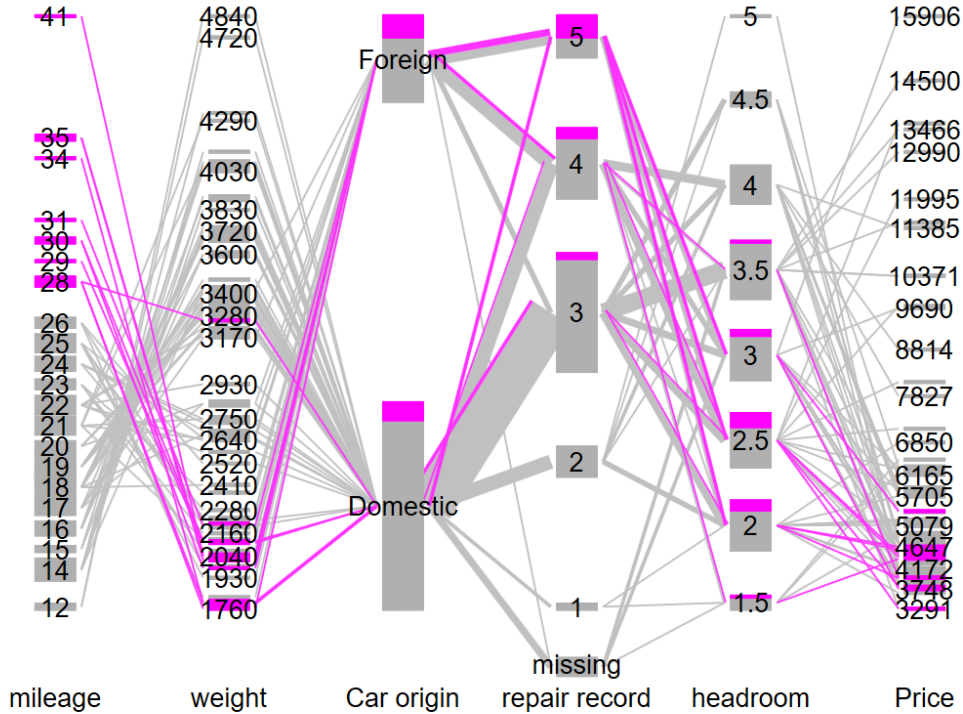}
\caption{Hammock plot of the auto data. Cars with high mileage are highlighted. The text explains how to choose colors rather than rely on default colors.}
\label{f:auto_mpg}
\end{figure}

Rather than relying on default colors, we can specify  colors in the option \texttt{colorlist}. 
Figure~\ref{f:auto_mpg}) shows the auto data where cars with a high gas mileage are highlighted (specifically, \texttt{mileage>27}).
We see  cars with high mileage are cheaper, they tend to have lower weight, a better repair record (rep78), and less headroom.

Figure~\ref{f:auto_mpg}  was created with the following command: 

\begin{stlog}
. hammock mpg weight foreign rep78  headroom price, ///
> missing ///
> hivar(mpg) hival(>27) ///
> label_format(\%8.0g) ///
> colorlist(gs11 sandb)
\nullskip
\end{stlog}
The first color in option \texttt{colorlist} is the default color;  any subsequence colors are used for highlighting in the order specified in \texttt{hival()}. In the grayscale version of this Figure, the colorlist is replaced with \texttt{colorlist(gs12 gs8)}. You can list all named colors in Stata with \texttt{graph query colorstyle}; it is also possible to specify colors in RGB format.

We can also include colors created from the user-written \texttt{colorpalette} command \citep{jann2023color}. The \texttt{colorpalette}  command saves colors in RGB format (e.g., "0 0 255") in  \texttt{r(p)} . We can input this directly into the \texttt{colorlist} option

\texttt{. colorpalette lin vegetables}

\texttt{. hammock x1 x2 x3, hivar(type) hival(1 2 3) colorlist(`"  `=r(p)'  "')}

\subsubsection{Specifying Labels}
Figure~\ref{f:auto_mpg}) used default labeling: add labels for all values unless there are too many ($>8$). In that case, show only labels for the minimum and maximum.

We can change the default behavior either by changing what number of values is ``too many'' in option 
\texttt{label$\_$too$\_$many}, 
or by specifying individual labels for each axis. For the 6 variables in Figure~\ref{f:auto_mpg}, we could specify the labeling method for each variable using \texttt{label$\_$methodlist(none minmax all all all  minmax)}. This would remove the labels of the first variable, display labels for minimum and maximum for the second and sixth variable, and displaying all labels for the remaining 3 variables.


\subsection{Tracing individual observations\label{s:tracing}}
\begin{figure}[htbp]
\centering
\hspace*{-2cm}
\includegraphics[width=1.25\textwidth]{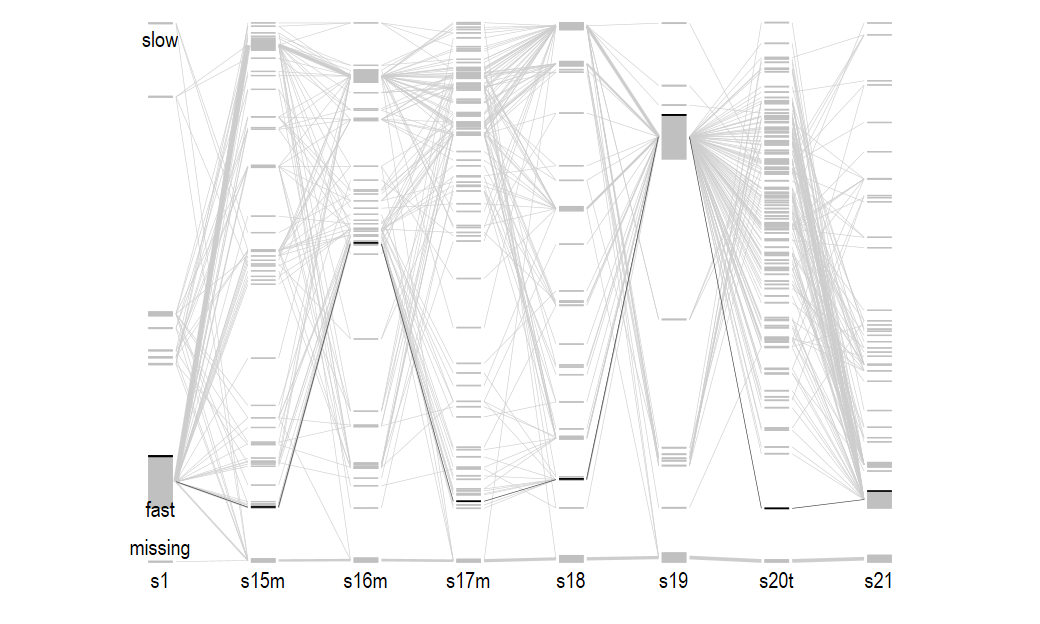}
\caption{Hammock plot of the first and the last stages of the tour de France data.  The eventual winner, Pogacar, is highlighted.  Each axis is scaled  individually from fastest to slowest rider. Times are not comparable across axes.}
\label{f:tdf_winners}
\end{figure}

Highlighting can also be used to trace individual observations across variables.  
Figure~\ref{f:tdf_winners} shows the finishing time for the first and the last several stages of the 176 riders of the  2020 Tour de France (see Appendix~A for details). 
The Tour de France is a bicycle race --- arguably the best-known bicycle race --- and is completed in 21 stages. Each stage corresponds to one racing day. Riders that finish as a group (without a visible gap of 1 second or at least one bike length) receive the same finishing time.
At the very bottom of Figure~\ref{f:tdf_winners} for each axis, there are missing values (riders who quit the race). Near the bottom, above the missing values, is the fastest stage time for that day, and at the top the slowest stage time for that day.  

The main group of riders is called the peloton. In bicycling, it is advantageous to ride in groups to draft off other riders. You can see the peloton across the stages. Additional groups may form as riders try to break out ahead of the peloton (break-away groups and chase-groups) or cannot keep up (the so-called ``autobus'' group). In particular, in the mountain stages several groups may form. Mountain stages are those with variable labels ending in an ``m''. 

Stage 20 is a time trial indicated by the variable label  ending with a ``t''.  
In a time trial,  riders ride individually and cannot draft off their teammates. There are no groups in the time trial. 

Looking at the riders with ``missing'' finishing times, i.e.,  riders who  had to quit the race for one reason or another, we find a surprise: 
In stage 17, there is a line leading out of the missing group, suggesting a rider did not finish stage 17 but re-entered the race in stage 18. This rider is Bryan Coquard and upon investigation it turned out that this is an error at the data source. We did not correct this here to showcase the plot's ability to find such errors.

We can trace the eventual winner, Pogacar, by highlighting his stage times. Pogacar was not always in the leading group. In stage 16 and 19 he let a number of riders pass. Pogacar did have the fastest time trial in stage 20 though.

Figure~\ref{f:tdf_winners} was created as follows:

\begin{stlog}
.         hammock s1  s15 s16 s17 s18 s19 s20 s21, ///
>         missing ///
>         nolabel ///
>         colorlist(gs10 red)   ///
>         barwidth(.2) minbarfreq(5) ///
>         uni_fraction(.15) ///
>         text(97 1 "slow" 10 1 "fast"  3 1 "missing") ///
>         hivar(rider) hival("Pogacar") 
{\smallskip}
\nullskip
\end{stlog}
We removed the labels on the axes (\texttt{nolabel}) to avoid clutter. \texttt{barwidth(.2)} reduces the width of the connectors (relative to the default \texttt{barwidth(1)}). Sometimes connectors of individual observations are barely visible because the connector is too thin. \texttt{minbarfreq(5)} increases the width of any connectors with fewer than 5 observations as if they had 5 observations. That is, connectors with 5 or fewer observations all have the same width. Also note the variable used for highlighting, rider, is not among the variables plotted. 
\texttt{uni\_fraction(.15)} scales the bars on the axes to cover 15\% of the available vertical space. The remaining 85\% is white.
This option can be adjusted to avoid overlap between the bars.  Some text is added (``slow'', ``fast'', ``missing''). The text is placed with y-values between 0-100 (here: 97, 10, 3) and with the x-value 1 corresponding to the first axis.

\begin{figure}[htbp]
\centering
\hspace*{-2cm}
\includegraphics[width=1.25\textwidth]{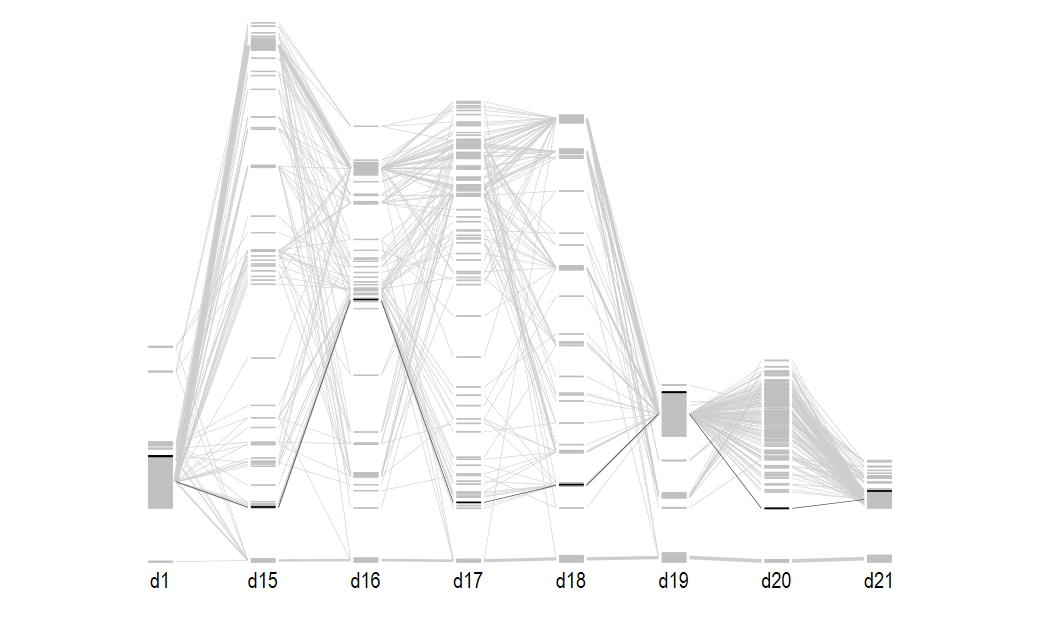}
\caption{Hammock plot of time lost to stage winners for multiple stages. The higher a rider on the plot, the more time he lost. All axes are on the same scale. The eventufual winner, Pogacar, is  highlighted.}
\label{f:tourdefrance_winners_relative}
\end{figure}

Figure~\ref{f:tdf_winners} nicely shows fast and slow riders. However, a unit of time is shown differently  on each parallel axis: the difference between the slowest time (at the top) and the fastest time (at the bottom, above the missing values) is not constant across stages; some stages are much longer than others. 
We subtract the time of the fastest rider for each stage and denote the new variables d1-d21. These new variables represent  how many seconds have passed since the stage winner crossed the finishing line. 
We can then use the option \texttt{samescale(varlist)} to enforce the same scale for all variables.
The revised plot is shown in Figure~\ref{f:tourdefrance_winners_relative}.

We see that riders lost a lot more time in stages 15-18 relative to the stage winner. There was little action on day 19: most riders finished in the peloton.
On the last day, day 21, it is customary for all riders to ride as a group except for the sprint at the end. This is why stage 21 shows one big peloton with other riders a few seconds behind. 

For the revised plot, Figure~\ref{f:tourdefrance_winners_relative} , we first computed the d1-d21 variables by subtracting the fastest stage time. The new variables are named d`i' (rather than s`i') with `i' ranging from 1 to 21.  The variables are created first:

\begin{stlog}
.         foreach var of varlist s* {\lbr}
  2.           quietly summarize `var', meanonly
  3.           local newname = "d" + substr("`var'", 2, .)
  4.           quietly gen `newname' = `var' - r(min)
  5.         {\rbr}
{\smallskip}
\nullskip
\end{stlog}

The hammock command follows: 

\begin{stlog}
.         hammock d1 d15 d16 d17 d18 d19 d20 d21, ///
>           nolabel ///
>           missing /// 
>           colorlist(gs10 red)   ///
>           barwidth(.2) minbarfreq(5) ///
>           uni_fraction(.15) ///
>           hivar(rider) hival("Pogacar") ///
>           samescale(d1 d15 d16 d17 d18 d19 d20 d21) 
{\smallskip}
\nullskip
\end{stlog}
\clearpage
\subsection{A connectors-only Hammock plot\label{s:connectors}}
A hammock plot consists of alternating displays of univariate statistics (bars) and bivariate statistics (connectors). When concentrating on the transitions, it sometimes makes sense to only show the connectors.

\begin{wrapfigure}{r}{0.50\textwidth}
  \centering
  \vspace{-10pt}
  \includegraphics[width=0.50\textwidth]{\figfile{stata_output/gesis_marital2}}
  \vspace{-10pt}
  \caption{Marital status over  5 waves of a survey panel. This hammock plot is identical to  Figure~\ref{f:gesis_marital} except that here univariate bars are included.}
  \label{f:gesis_marital2}
\end{wrapfigure}
\begin{figure}[htbp]
\centering
\includegraphics[scale=.35]{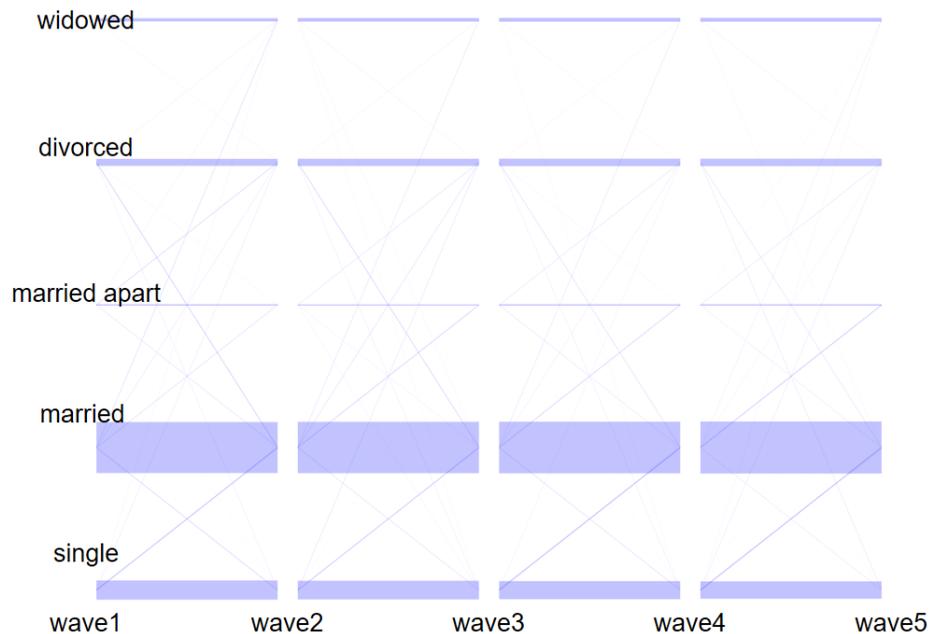}
\caption{Marital status over 5 waves (or time points) in the GESIS panel. Surprisingly, some married individuals  return to ``single'' rather than to ``divorced'' (or similar)   the following year.}
\label{f:gesis_marital}
\end{figure}

 Figure~\ref{f:gesis_marital2} shows 
marital status over 5 time points (or waves) in the GESIS panel (\url{https://www.gesis.org/en/gesis-panel}), a survey panel in Germany.  A panel survey is a longitudinal survey where the same individuals (or units) are surveyed repeatedly over time. Here, both the bars and the connectors are shown. 
The most common category in Figure~\ref{f:gesis_marital} is ``married'' (living together rather than apart), followed by ``single''. 
We can see that marital status does not change much within a few short years. The connectors look very similar to the univariate bars; the univariate bars take up space without adding information. 

Figure~\ref{f:gesis_marital} shows the same  plot but removes the univariate bars.
Now we can more easily concentrate on the transitions from wave to wave. What might surprise is the small subset of observations that went from ``married'' to ``single'' (rather than ``divorced'' or one of the other categories). It turns out that marital status is self-assessed  and some respondent preferred ``single'' over ``divorced'' or another category.

Figure~\ref{f:gesis_marital} was created as follows: 

\begin{stlog}
. hammock wave1 wave2 wave3 wave4 wave5, ///
>   nolabel /// 
>   nounibar ///
>   barwidth(3) ///
>   space(0.1) ///
>   colorlist(blue\%30) ///
>   minbar(20) ///
>   text(16 1 "single" ///
>      38 1 "married " ///
>      57 1 "married apart" ///
>      80 1 "divorced" ///
>     100 1 "widowed" ) ///
>   graphregion(margin(l+10 r+10))
\nullskip
\end{stlog}

As before, the grayscale version of this Figure can be created by replacing \newline 
\texttt{colorlist(blue\%30)} with  \texttt{colorlist(gs8)}. \texttt{nounibar} requests not to plot univariate bars on the axes.  
\texttt{space(0.1)} leaves some space between the connectors.
We specified \texttt{nolabel} because the same label applies for all variables. Instead,
we added the labels by hand using Stata's  
\texttt{text} option which is passed through to the plotting command. \texttt{minbar(20)} requests that each connector has a minimum thickness (the minimum thickness is specified in terms of the number of respondents it represents, here 20 respondents). The x-coordinate for the text is always $1$ because the text is for the first x-variable and  the first x-variable has the x-coordinate $1$. The y-coordinates are spaced between 0  and 100; the exact values are determined by trial and error. (Temporary optional arguments such as \texttt{yline(10 20 30 40 50)} can help finding desired coordinates). 
\texttt{barwidth(3)}  triples the width of the connectors.
We add some space on the left and right using Stata's \texttt{graphregion(margin())} option. This avoids that some text runs off the page.
For variety, we specified a different color using a color intensity of 30\%. We did not reserve an area for missing values in this plot. 

\subsubsection{More on the size and shape of bars and connectors}
\texttt{uni\_fraction(0.5)} controls the height of the univariate bars. By default, 50\% of the vertical space is filled with univariate bars. If any two bars overlap, this fraction can be adjusted downward to avoid the overlap. This fraction can also be adjusted upward to suit.
Likewise, \texttt{barwidth(1)} can be adjusted to control the width of the connectors to reduce visual clutter or to reveal more detail. The number specified is an adjustment factor relative to the default, 1 (see an example in Figure~\ref{f:gesis_marital}).
connectors that represent a very small fraction of observations can be so thin that they are barely visible or do not render properly. 
As seen in the code for Figure~\ref{f:tdf_winners}, option \texttt{minbarfreq(1)} specifies that a connector must represent at least \texttt{minbarfreq} observations.  All bars corresponding to fewer than \texttt{minbarfreq} observations are displayed as if they contained \texttt{minbarfreq} observations. The default value, 1, does nothing because all connectors  contain at least 1 observation. 

The option \texttt{shape(rectangle)} controls the shape of the connectors: rectangle or parallelogram. The main advantage of the parallelogram is computational speed. The rectangle shape fixes the so-called reverse line width illusion (see \cite{schonlau2024hammock}). The helpfile describes additional layout options. 
\subsection{Parallel univariate plots\label{s:paruni}}
\begin{figure}[htbp]
\centering
\hspace*{-2cm}
\includegraphics[width=1.30\textwidth]{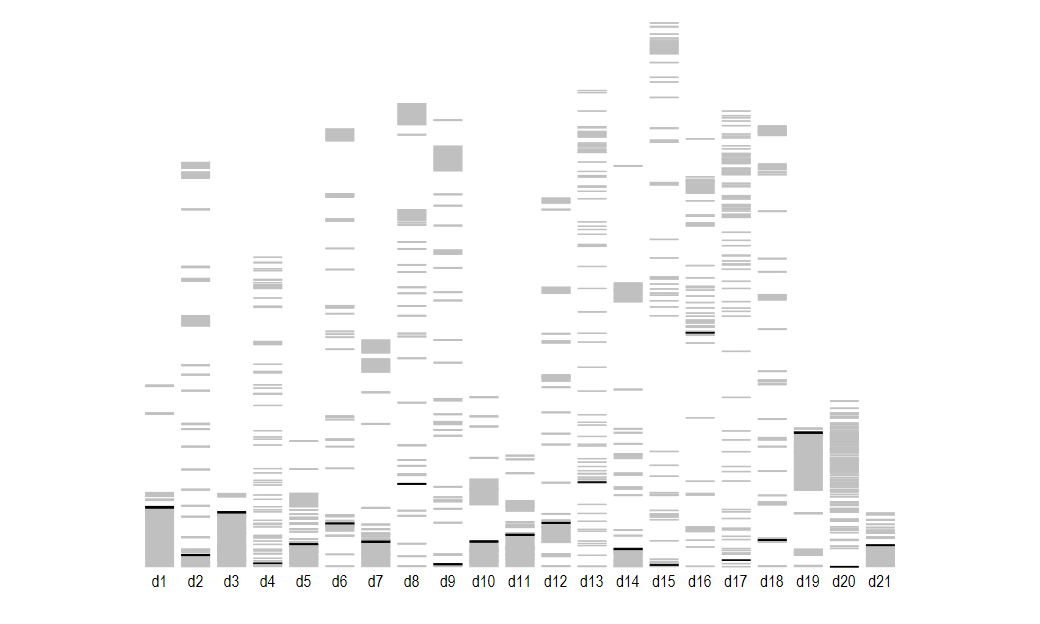}
\caption{Parallel univariate plot of the 2025 Tour de France. The winner of the tour, Pogacar, is highlighted. Except in stage 16, Pogacar was never far behind the stage winners.}
\label{f:tourdefrance_paruni}
\end{figure}

A hammock plot shows an alternating sequence of univariate plots (univariate bars on the axes and bivariate plots (connectors between two neighboring axes). If we omit the bivariate plots, parallel univariate plots emerge as an edge case. In Stata, we simply specify \texttt{space(1)}.  Figure~\ref{f:tourdefrance_paruni}  shows an example for the 2020 Tour de France data. Without the connectors we  are able to display more variables.
We see that the winner, Pogacar, was not always in the winning group, but except for stage 16 he was never far behind.

The parallel univariate plots are connected with one another through highlighting. 
The plot was constructed with the following command:

\begin{stlog}
.         hammock d*, ///
>           nolabel ///
>           colorlist(gs10 red)   ///
>           barwidth(.2) minbarfreq(5) ///
>           uni_fraction(.15) ///
>           hivar(rider) hival("Pogacar") ///
>           samescale(d*) /// 
>           space(1) 
{\smallskip}
\nullskip
\end{stlog}

\begin{figure}[htbp]
\centering
\hspace*{-2cm}
\includegraphics[width=1.25\textwidth]{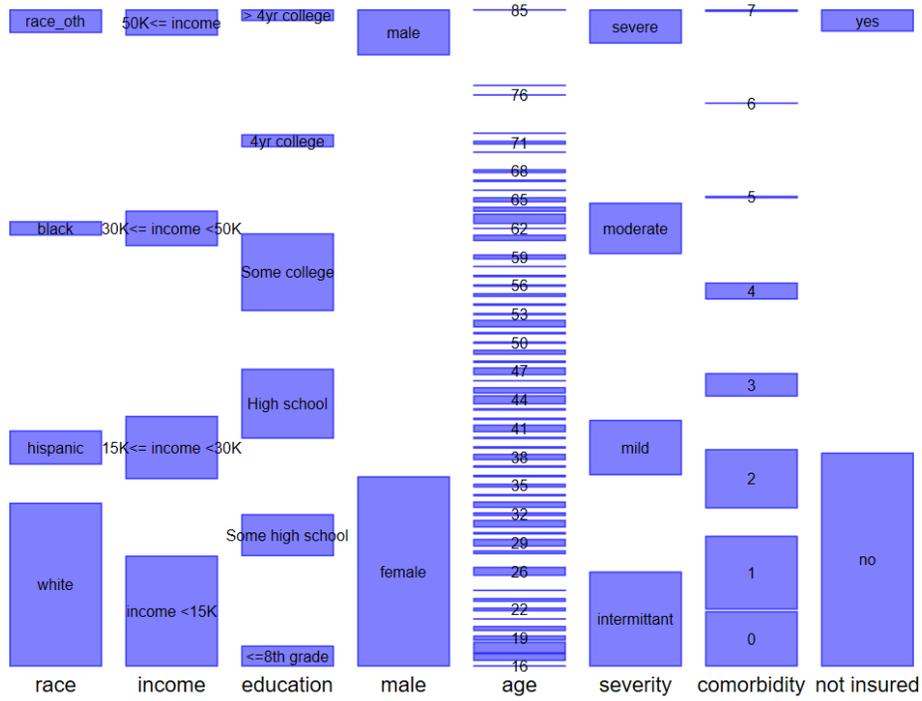}
\caption{Parallel univariate plot of the asthma data. The same data are shown as a table in Table~\ref{t:asthma}. }
\label{f:asthma_uni}
\end{figure}

In empirical studies, ``Table 1'' often shows the demographic composition of the  study participants. Such a ``Table 1'' is shown in Table~\ref{t:asthma} with percentages for categorical variables and mean/ standard error for continuous ones. The data are from an asthma study \citep{schonlau2005evaluation}.  Alternatively, one might show a parallel univariate chart plot as shown in Figure~\ref{f:asthma_uni}. We see a graphical representation of separate demographic x-variables.
The univariate charts not linked in any way. However, we could easily highlight a specific category, such as female participants, and observe the percentage of female participants in each category.

\begin{table}[htbpp]
    \centering
    \begin{tabular}{llr}
        \toprule
        Variable & Category &  Percent  \\
        \midrule
    Race  & black & 6.5 \\
         & hispanic & 14.6 \\
          & white & 69.2 \\
         & other race & 9.7 \\
    Income  & $<$ 15K & 47.2 \\
          &  15K $\le$ income$<$ 30K & 26.7 \\
          &  30K $\le$ income$<$ 50K & 15 \\
           &  50K $\le$ income & 11.1 \\
    Education   & $<=$ 8th grade & 9.2 \\
          &  some high school & 17.8 \\
          &  high school & 29.2 \\
           &  some college  & 32.4 \\
           &  4yr college & 6.0 \\
           &  $>=$ 4yr college  & 5.4 \\
                  Gender  & male & 20.0 \\
         & female & 80.0 \\
    Age   & mean  & 42.8 \\
         & SD  & 16.8 \\
        Severity  & intermittent & 40.5 \\
         & mild & 23.8 \\
          & moderate & 21.1 \\
         & severe & 14.6 \\
    Comorbidities   & mean  & 1.57 \\
        & SD & 1.38  \\
    Insurance   & no  &  90.8 \\
         & yes  & 9.2  \\
        \bottomrule
    \end{tabular}
    \caption{Demographic information about the study population in a study related to asthma. The same data are shown in  parallel univariate plots in Figure~\ref{f:asthma_uni}.}
    \label{t:asthma}
\end{table}

The plot corresponding to Table~\ref{t:asthma} was constructed with the following command:

\begin{stlog}
. hammock race income education male age severity comorb no_ins, ///
>   space(1) ///
>   uni_fraction(.36) ///
>   colorlist(orange\%20) ///
>   labelopt(size(vsmall))  ///
>   xlab(,labsize(small)) 
\nullskip
\end{stlog}
The option \texttt{labelopt(size(vsmall)} reduces the font size of the labels on the axes; the option \texttt{xlab(,labsize(small)} reduces the font size of the variable names. 
The grayscale version of this plot uses the option  \texttt{colorlist(gs12)}. The value for option uni$\_$fraction (which affects the relative size of the univariate bars) was chosen as large as possible without the two univariate bars corresponding to 0 and 1 comorbidities touching. (Spacing for numerical variables is not arbitrary: If one were to move a univariate bar for comorbidities  upwards, then the variable would become categorical because the meaning of distance is lost.)


\subsection{Re-ordering variables and categories}
The variable order can be rearranged to make the overall plot more understandable or to make a specific point.
As a starting point, one may arrange the order to maximize the squared correlation of neighboring variables (or another measure of association for categorical variables). For the auto data, Figure~\ref{f:auto_foreign} shows a variable ordering  starting with mileage (on the left), where the following variable is chosen to maximize the squared correlation with the current variable.
For example, among the remaining variables, weight has the highest squared correlation with mileage, ``car origin'' has the highest squared correlation with weight, and so forth.

Figure~\ref{f:auto_foreign} shows mileage and weight are negatively correlated; domestic cars are heavier than foreign cars, and the repair record of foreign cars is better. The relationship with price is not obvious, and we might start to highlight expensive cars. 

\begin{figure}[htbp]
\centering
\includegraphics[width=1\textwidth]{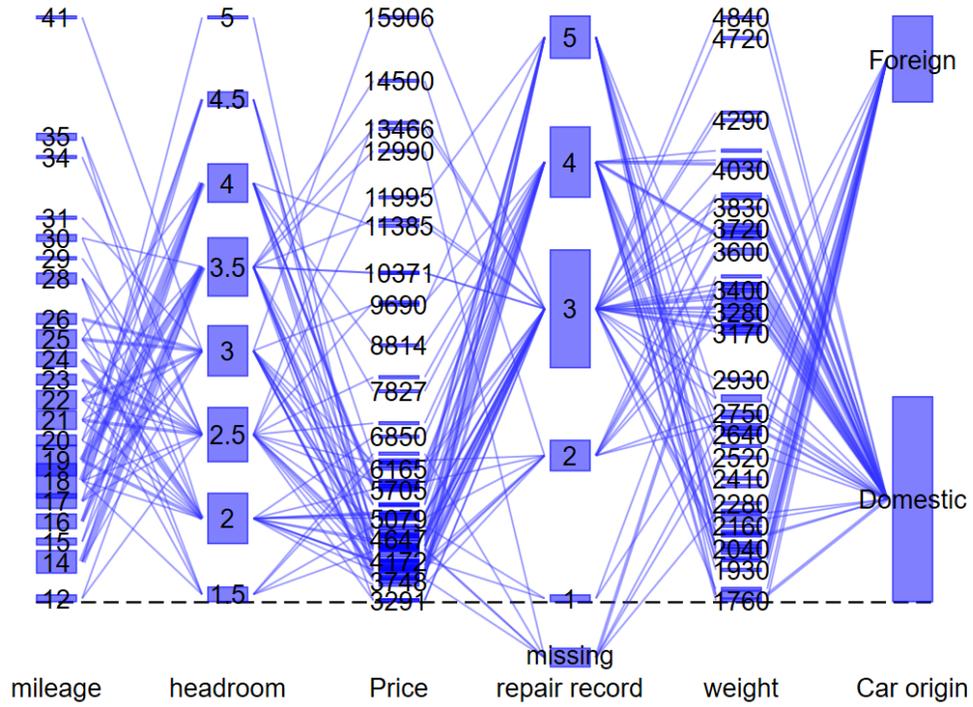}
\caption{Hammock plot of the auto data: variables are ordered to minimize the squared  correlation among neighboring variables. Instead, we prefer the ordering in Figure~\ref{f:auto_foreign}.}
\label{f:auto_order_low}
\end{figure}

To better understand why the ordering in Figure~\ref{f:auto_foreign} may be a sensible starting point, we now do the opposite: again starting with mileage, we continue with the variable with the lowest squared correlation to mileage, and so forth. The resulting plot is in Figure~\ref{f:auto_order_low}. 
We believe that this Figure is somewhat harder to read. The relationship among the first five variables is less obvious. We can still see that foreign cars tend to weigh less as these two variables remain in the same order.

\subsubsection{Re-arranging categories}
When the variable is continuous or ordered categorical, the order of the values and categories is fixed (in rare circumstances it may be sensible to reverse the order). When a variable is unordered categorical, we can re-arrange the categories to make the graph more readable or to bring out an aspect of the story we wish to tell. Often, a graph is easier to read when  ---  among neighboring variables --- low univariate bars connect to low univariate bars and high univariate bars connect to high univariate bars. 

The only unordered categorical variable in Figure~\ref{f:auto_foreign} is car origin (foreign vs domestic). In the current order, connectors from domestic (at the bottom) connect to low repair univariate bars (at the bottom). This accentuates the finding that domestic cars have worse repair records.

\section{\label{s:conclusion}Conclusions}
The hammock plot visualizes categorical and numerical variables within a single, unified plot. By mapping categories to parallel vertical axes and connecting observations across variables, the hammock plot supports simultaneous inspection of distributions, relationships, and individual trajectories. This is useful for both analysis and communication and makes hammock plots applicable in a wide range of settings.

In exploratory data analysis, hammock plots enable analysts to quickly assess relationships among multiple variables. Without highlighting, univariate and select bivariate relations (those of neighboring variables) are shown. With highlighting, additional bivariate and select tri-variate relationships are also shown.
 There is a dual focus on distributions and highlighting individual cases; and this distinguishes hammock plots from traditional summary graphics such as bar charts or boxplots.

Hammock plots are effective for the visual identification of outliers. Unlike numerical detection methods, this visual approach allows analysts to judge outliers in the context of multiple variables simultaneously, which is particularly useful when deviations are meaningful only in relation to several variables.

Hammock plots are useful in uncovering missing value patterns. At a glance, analysts see the extent of missingness for each variable.  Highlighting missing values of one variable, analysts can determine, for example, whether missingness across variables is mostly  due to the same subset of observations. If not, such highlighting may reveal a more intricate missing value pattern such as a dependency on another variable.

Hammock plots are  well suited for comparative analysis. Highlighting a group of observations shows whether the observations are  systematically different than the remainder across a range of variables.  For example, highlighting is useful to compare outcomes for a treatment versus a control group.

Hammock plots support tracing individual observations across multiple variables. This is particularly intuitive for  longitudinal data or when the sequence of variable is ordered in time or otherwise meaningful.  This capability supports visual story telling and helps connect statistical summaries to concrete cases.

Hammock plots give a visual summary of the data and can serve as an alternative to tables. While tables provide exact values, hammock plots visually distinguish categories of different sizes  and can showcase relationships. This makes them especially useful when communicating findings to audiences interested in patterns rather than precise numerical detail.

 Among the variations of the hammock plot is the alluvial plot, developed later and unable to accommodate numerical variables.
The hammock plot (or one of its variations) is not the only way to gain insight into data.  It is, however,  one more tool in your toolbox and it is often useful.

In summary, hammock plots are a flexible visualization technique that bridges summary-level insight and observation-level detail, supporting both exploration and communication for multiple mixed categorical/numerical variables.

\section*{\label{s:appendix}Appendix A: 2020 Tour de France data}
The 2020 Tour de France data is  made available as part of this paper. They were scraped from the website 
 \url{https://www.bikeraceinfo.com/tdf/tdf2020.html}  by the authors. 
The data contain rider name, team name, 21 variables (s1-s21) for the stage finishing times, and 21 variables for the cumulative finishing times (c1-c21). All finishing times are given in seconds. Variable labels (not: variable names) of mountain stages and time trials are denoted with ``m'' and ``t'', respectively. 

Cumulative  times were also scraped from the website listed above. They differ from the sum of the individual times: for the 2020 Tour de France, the first/second/third place winner of any stage was awarded a reduction of 10/6/4 seconds to their overall time.

 As pointed out earlier, the entry for Stage 17 for Bryan Coquard is missing.  Based on the website 
 \url{https://www.procyclingstats.com}, 
 he finished  35min 45sec  after the stage winner.  His value for s17 should be 2145 (seconds).  
 
\bibliographystyle{sj}
\bibliography{hammock}

\begin{aboutauthors}
Matthias Schonlau is a professor of statistics in the department of Statistics and Actuarial Science at the University of Waterloo, Canada.
His interests include visualization and survey methodology, in particular the analysis of open-ended questions.

Tiancheng Yang is a Ph.D. student in the department of Statistics and Actuarial Science at the University of Waterloo, Canada. His research focuses on machine learning, natural language processing and visualization tools for survey and social science applications.
\end{aboutauthors}

\clearpage
\end{document}